\documentclass[a4paper,12pt]{article}
\usepackage{graphicx}

 \author{Munir H. K. Al-Hashimi\\
Institute of Theoretical Physics, Bern University\\ Sidlerstrasse  5, 3012 Bern, Switzerland}
  \title {Discrete Momentum Mechanics and Faster Than Light Transition}

 \date{2 February 2004}
\begin{document}
\maketitle
 \pagestyle{myheadings}\markright{Munir H. K. Al-Hashimi}
 \begin{abstract} In this work a new
mechanics will be studied which is based on the hypothesis that
the change of linear momentum of a particle happens as a discrete
pulses. By using this hypothesis and by considering Newton's
relation between energy and momentum, and the law of mass and
energy conservation as a priori, the Einstein dispersion relation
can be derived as a zero approximation without using Lorentz
transformations. Other terms will be derived as a corrections to
this relation. It will be shown that the effect of the corrections
will be smaller and smaller with the increase of momentum. The
work will offer an explanation of why the velocity of light seems
to be constant regardless of the velocity of the source, and under
which condition this will be changed. Also a prediction is made
that faster than light transition could happen theoretically under
certain conditions, and a nonzero mass photon can exist in nature.
The work is purely classical in the sense that it doesn't involve
any uncertainty relations.
\end{abstract}

 \section{Introduction}
One of the unanswered questions in modern physics is whether the
photon has a  tiny rest mass or not. From a classical point of
view and without the notion of the theory of relativity it is
totaly unacceptable to assume that a particle like the photon has
a zero rest mass. Such a suggestion would be rejected for the
following reasons:
\begin{enumerate}
  \item
   How come a particle that is
 carrying energy like any other particles does
 not have something which the rest of the particles have in common, that is a
 rest mass.
 \item
 From the Newtonian Mechanical point of view, zero mass means that a particle will be
 accelerated to infinite velocity by any force no matter
 how small it is.
\item
  If a particle reaches a certain velocity it must be accelerated from rest to this velocity
 ; it meaningless to say that a particle has no mass at rest, or,
 in other words has no existence at rest, and then when this nonexisting entity gain
 momentum it came to creation. Again this may be acceptable from a quantum
 a mechanical point of view, but not from classical point of view.
 \end{enumerate}
 Even after Special Relativity there are scientists who thought
 that the photon may have a very small rest mass. To shed light
 on this point I will quote from a conversation between Feynman
  and a French professor [1]. The professor asked ``Tell me Professor Feynman,
  how sure are you that the
  photon has no rest mass,?". Feynman answered``Well, it depends on
  the mass; evidently if the mass is infinitesimally small, so
  that it would have no effect whatsoever, I could not disprove
  its existence, but I would be glad to discuss the possibility
  that the mass is not certain definite size. The condition is
  that after I give you arguments against such mass" The professor then chose
  a mass of $10^{-6}$ electron mass, Feynman answered``if we agreed
  that the mass of the photon was related to the frequency
  as $\omega=\sqrt{k^2+m^2}$ , photons of different wave lengths would
  travel with different velocities. Then in observing the eclipsing double star
  ,which is sufficiently far away, we would observe the eclipse in
  blue light and red light at different times. Since nothing like this is
   observed we can put an upper limit on the mass, which, if
  we do the numbers, turns out to be of the order of
  $10^{-9}$ electron mass". The professor asked if it is possible that the
  photon has a mass of $10^{-12}$. Feynman answered``If the photon
  has a small mass equal for all photons, larger fractional
  differences from the massless behavior are expected as the wave
  length gets longer. So that from the sharpness of the known
  reflection of radar pulses, we can put an upper limit to the
  photon mass which is somewhat better than the eclipsing double
  star argument. It is turn out to be smaller than $10^{-15}$ electron mass". After this the professor
  asked what if the mass $10^{-18}$ electron mass. Feynman answered
  ``From field theory argument the potential should go as
   $exp(-mr)/r$. Then the earth has a static magnetic field, which
  is known to extend out into space for some distance, from the
  behavior of the cosmic ray of the order of few earth radii. But
  this means that the photon mass must be of a size smaller than
  that corresponding to a decay length of the order of 8000 miles,
  or some, $10^{-20}$ mass". \par But what if the mass of the photon is
  much smaller than that. In fact  L. de Broglie noted that a rest mass a photon
 of order $10^{-65}$ g would be impossible to
  detect.
 \par Here it must be noted that, if there is really a photon with a rest mass no matter
 how small it is, then this will impose a change on the basic assumptions of
 the theory of special relativity because the second postulate
 stated that the velocity of light is always constant and it is
 equal to $c$ no matter what is the velocity of the source [2] . Clearly
 that is not the case for a photon with a rest mass, because the
 velocity of light will depend on the velocity of the source as
 well as the light's wave length. The second postulate only
 applied to photons with infinity energy since only then the
 velocity of light will be independent of the velocity of the
 source and it will be exactly $c$ , and practically there is no such photons.

\subsection{The Purpose and Outline of this Work} The purpose of this work is to
introduce a new scheme under which the relativistic energy
momentum relation (Einstein Dispersion Relation) can be derived
without using Lorentz transformations and without putting any
limit on the maximum velocity with which particles can propagate,
and to investigate the possibility of rest mass photons.
\par The treatment is based on the assumption that the exchange of
momentum between particles is not a continuous process, but it
happens in the form of discrete pulses. The derived dispersion
relation is an approximation to Einstein dispersion relation (EDR)
; an approximation that is becoming very
 high with the increase of energies. It will be shown that under such an assumption it
 is possible for particles with nonzero mass to travel at the
 velocity $c$ and faster, which means that it is possible for the photon to have a nonzero rest mass. It will also be shown
 that the new understanding will explain
 the results of the Michelson-Morley experiment. The constancy of the velocity of
 light will be a result and not a postulate like in the case of SR. However it will be shown
 that the principal of the constancy of the velocity of light can be violated for low energy photons.
 \par The second section will discuss the basic postulates of the
 work, the third and fourth will show the steps to find the total energy
 of a particle and the correction terms to EDR as a power
 series. Section four will show a way to find the correction as
 functions of momentum. Section five will examine the convergence of the solution,
 section six will discuss the relation between velocity and
 momentum and between velocity and energy, and faster than light
 transition, the last section will be for conclusions.
\section{The Basic Postulates}
In this work there are two postulates and one preposition.
 \subsection{First Assumption} Particles in nature  are interacting through exchanging
 momentum. We are use to deal with momentum in classical mechanics and in most of
 the cases of quantum mechanics as a continuous quantity. The core
 idea of this work is to deal with linear momentum as a sum of
 vectors of small values each of which has a universal magnitude denoted by $p_s$,
 this make linear momentum in some sense
 a discrete vector entity. The first postulate stated that:
 \emph {I- The change of the linear momentum of a body or
     a particle with time has a minimal vector quantity of
    $\overrightarrow{p}_{s}$ , with $p_s$ as a universal constant}.
    To understand what exactly this postulate means let us consider the change of
    momentum, say from zero value to a value
    $\overrightarrow{p}$ during a time $\Delta t$.
    The interval $\Delta t $ can be divided in to smallest possible
    subintervals $\Delta t_{1},\Delta t_{2}, ...,\Delta t_{i}$ such that in each of
    them the momentum is changing
     by  the smallest possible value. Since according to the first postulate, the
     smallest possible change is   $\overrightarrow{p}_{s}$, the
      momentum of a particle can be written as
\begin{equation}
  \overrightarrow{p}=\sum_{l=1}^{i}\overrightarrow{p}_{sl}
\end{equation}
, where $\overrightarrow{p_{s1}}$ is the change of momentum during
$\Delta t_{1}$.  $\overrightarrow{p_{s2}}$ is the change of
momentum during $\Delta t_{2}$ ,...and so forth, and
\begin{equation}
  |\overrightarrow{p_{s1}}|=  |\overrightarrow{p_{s2}}|=...=p_s
\end{equation}
 It is assumed here that the particle was initially at rest. Since
the discrete nature of linear momentum is so far undetectable
experimentally, this suggests that the value of $p_{s}$
 is extremely small. It is so small that the momentum of the
 particles seems to be continuous.
\subsection{ Second Postulate}The equivalence between energy and
mass can be reached without using relativity. In fact, a work by
Poincar\'{e} [3] had led to such a relation by studying the
momentum of radiation. Another work by Hasen\"{o}hrl [4],[5] can
lead to the relation by studying the pressure inside a system
composed of a hollow enclosure filled with radiation. Braunbeck
[6] had shown in 1937 that the verification of the mass energy
equivalence relation by experiment must not be regarded as a
theorem that can be derived from other principles of less direct
and less empirical evidence, but should be taken as a fundamental
principle. This point of view will be adopted here. The mass
energy relation in this work is a postulate (or a principle), but
I stress here that in the mass energy relation.
\begin{equation}
  E=mc^2
\end{equation}
, the constant $c$ will not be defined as the velocity of light,
 it will be considered only as a constant that has the units of
velocity. It will be one of the results of this work that
particles with $p\gg m_oc$ will travel with a velocity approaching
$c$. The second postulate is: \emph{mass and energy are
equivalent.}
 \subsection{ Phenomenological preposition }
\par The third postulate is supposed to specify the relation between the energy change
 of a particle due to a vector change of one momentum pulse $\overrightarrow{p_s}$  and the momentum vector
  $\overrightarrow{p}$. By knowing this relation, the dispersion relation can
 be found after summing the values of energies due to a change of
 momentum by
 $i$-pulses.
 \par To show how this phenomenological relation has been suggested, consider
 a particle that has a change of linear momentum equal to
  $\overrightarrow{p}_{s}$. According to Newton's mechanics the
  change of energy due to this change of momentum will be
\begin{equation}\Delta
E= \frac{1}{m_{o}}\left(
\overrightarrow{p}_{bf}\cdot\overrightarrow{p}_{s}+\frac{p_{s}^{2}}{2}\right)
\end{equation}
where $\overrightarrow{p}_{bf}$ is the initial momentum vector of
the particle, or it  is the momentum vector before the change of
momentum by $\overrightarrow{p}_{s}$ , and $m_o$ is the rest mass
of the particle. In this work it will be purposed that  Newton
relation is applicable but with a difference, according to the
second postulate mass and energy are equivalent therefore the
kinetic mass must be used instead of the rest mass. More about the
preposition will be discussed in another work under preparation
[7]. Accordingly the change of the energy of a particle due to a
change of its momentum $\overrightarrow{p_{bf}}$ by one momentum
pulse is
\begin{equation}\Delta
E=\left(\frac{\overrightarrow{p_{bf}}\cdot\overrightarrow{p_{s}}}{m}+\frac{p_{s}^{2}}{2m}\right)
\end{equation}
\begin{equation}
 m= \frac{E^{(t)}}{c^2}
\end{equation}
where $E^{(t)}$ is the total energy of the particle. Relation (5)
will be used for calculating the energy that is gained by the
particle due to  the $i^{th}$ momentum pulse.
\par According to the previous discussion the state of the proposition will be:
\newline  \emph{ if a particle with momentum ${p_{bf}}$ changed its
momentum by one momentum pulse then the change of energy due to
that will be according to relation(5)}.
 \par In this work only an accelerated particle from rest will be
studied, it will be assumed that the acceleration process will not
cause any change in the inner state of particle, and therefore the
rest mass will not change. If a force acting on a particle has
fixed direction, then it would be reasonable to believe that all
the momentum pulses will be pointing in the same
 direction of the accelerating force. Accordingly, for all momentum pulses,  $\theta =0$ in relation
 (5).
The total linear momentum of the particle after the end of the
action of the accelerating force will be:
\begin{equation}
  p=ip_s
\end{equation}
, and therefore the energy change due to the $i^{th}$ momentum
pulse can be written in the following form
\begin{equation}\Delta
E_{i}=\left(\frac{(i-1)p_{s}^{2}c^2}{E^{(t)}_{i}}+\frac{p_{s}^{2}c^2}{2E^{(t)}_{i}}\right)=\left(\frac{(i-\frac{1}{2})p_{s}^{2}c^2}{E^{(t)}_{i}}\right)
\end{equation}
, where $ E^{(t)}_{i}$ is the total energy of the particle due to
$i$ momentum pulses, or
\begin{equation}
  E^{(t)}_{i}=m_{o}c^2+\sum_{l=1}^{i}\Delta E_{l}
\end{equation}
Here it must be noted that quantities like $p_s$, $m_o$, $E^{(t)}$
...etc. are measured relative to one fixed observer.

\section{Calculating The Total Kinetic Energy}
 The total kinetic energy $K$ of a particle can be calculated by
 adding the kinetic energy gained from the first pulse plus the
 energy gained from the second pulse ... and so forth. Therefore
 $K$ can be expressed as
\begin{equation}
  K(i)=\sum_{k=1}^{k=i}E_k
\end{equation}
, where the delta was dropped from $\Delta E_k$ for short hand.
\par Expressing $E_k$ will become more problematic with the
increasing of $k$. This can be shown by calculating $E_1$ and
$E_2$ for a particle with rest mass $m_o$ , for example, (8) will
give
\begin{equation}
  E_1=\frac{1}{2}\left(-m_oc^2\pm\sqrt{m_o^2c^4+2p_s^2 c^2}\right)
\end{equation}
The negative energy change solution will lead to a jump of the
total energy to negative value due to a small change in momentum
equal to $p_s$. The aspect of such solution will not be discussed
in this work. \par Here it is important to note that the positive
solution of (11) is deviating clearly from the SR solution, but
this will not contradict the experiment since there is no
measurement to energies of particles with low value of momentum
such as one or a few $p_s$. On the other hand, it will be shown
that $E^{(t)}$ will approach the one of SR as the momentum
increases.
\par The expression of $E_2$ will be found
by using the expression of $E_1$, because
$E^{(t)}=m_o\;c^2+E_1+E_2$. Therefore (8) and (11) give
\begin{eqnarray}
E_2=\frac{1}{2}\Bigg[-\frac{1}{2}\left(m_oc^2+\sqrt{m_o^2c^4+2p_s^2
c^2}\right)+\hspace{20mm}
\\\nonumber\sqrt{\frac{m_o^2c^4}{2}+\frac{13p_s^2c^2}{2}+\frac{m_oc^2}{2}\sqrt{m_o^2c^4+2p_s^2
c^2}}\Bigg]
\end{eqnarray}
It is clear that calculating $K$ in this way will be unpractical.
The fact that $p_s$ is very small will help in finding a series
solution so that the expression of $E_1$ , $E_2$,...,$E_i$ can be
written as a power series in $p_s$. For example, $E_1$ can be
expanded as
\begin{equation}
  E_1=\frac{1}{2}\frac{p_s^2}{m_o}-\frac{1}{4}\frac{p_s^4}{m_o^3c^2}+\frac{1}{4}\frac{p_s^6}{m_o^5c^4}-\frac{5}{16}\frac{p_s^8}{m_o^7c^6}+...
\end{equation}
, and  the expansion of $E_2$ will be
\begin{equation}
 E_2=\frac{3}{2}\frac{p_s^2}{m_o}-3\frac{p_s^4}{m_o^3c^2}+\frac{87}{8}\frac{p_s^6}{m_o^5c^4}-\frac{771}{16}\frac{p_s^8}{m_o^7c^6}+...
\end{equation}
The kinetic energy for this case, where $i=2$, is
\begin{equation}
  K(2)=E_1+E_2=2\frac{p_s^2}{m_o}-\frac{13}{4}\frac{p_s^4}{m_o^3c^2}+\frac{89}{8}\frac{p_s^6}{m_o^5c^4}-\frac{97}{2}\frac{p_s^8}{m_o^7c^6}+...
\end{equation}
For large $i$ this will also be an unpractical way of calculation.
A rule must be found for the coefficients of the power series for
$K$ for arbitrary values of $i$ . To reach to this rule $K$ will
be written as
\begin{equation}
  K(i)=\sum_{l=1}^{l=i}E_l=\sum_{j=1}^{j=\infty}g_j(i)\frac{p_s^{2j}}{m_o^{2j-1}c^{2j-2}}
\end{equation}
So the next step is to find $g_j(i)$ which represents the
coefficients of the expansion of $K$ for arbitrary $i$ . To do
that, (8) and (9) will be used to write the change in kinetic
energy due to the $i$ the pulse in terms of the total energy due
to $i-1$ pulses,
\begin{equation}
E_i=\frac{1}{2}\left(-(m_oc^2+\sum_{k=1}^{k=l-1}
E_k)+\sqrt{(m_oc^2+\sum_{k=1}^{k=l-1}E_k)^2+2p_{s}^2(2i-1)c^2}\right)
\end{equation}
Expanding $E_i$ as a power series in $p_s$ gives the following
expression
\begin{equation}
E_i=\frac{1}{2}\left[\sum_{j_0=1}^{\infty}
\frac{p_{s}^{2j_0}(2c)^{j_0}[2i-1]^{j_0}C(1/2,j_0)}{(m_oc^2+\sum_{k=1}^{i-1}E_{k})^{2j_0-1}}\right]
\end{equation}
, where the binomial theorem has been used to get the above
equation , and $C(1/2,j_0)$ are the binomial coefficients. Using
the binomial theorem again for the term in the denominator of (18)
gives
\begin{equation}
E_i=\frac{1}{2}\left[\sum_{j_0}^{\infty}
\frac{p_{s}^{2j_0}(2c^2)^{j_0}}{m_o^{2j_0-1}}[2i-1]^{j_0}C(1/2,j_0)\sum_{r=0}^{r=\infty}C(-2j_0+1,r)\left(\frac{\sum_{k=1}^{i-1}E_{k}}{m_o}\right)^r\right]
\end{equation}
From (16) it is easy to prove that
\begin{equation}\
(\sum_{k=1}^{i-1}E_{k})^r=\sum_{j_1=1}^{j_1=\infty}...\sum_{j_r=1}^{j_r=\infty}g_{j_{1}}(i-1)...g_{j_{r}}(i-1)\frac{p_s^{2(j_1+...j_r)}}{m_o^{2(j_1+...j_r)-r}c^{2(j_1+...j_r)-2r}}
\end{equation}
After substituting with
\begin{equation}
  j_1+...j_r=J
\end{equation}
, equation (20) can be written as
\begin{equation}
  (\sum_{k=1}^{i-1}E_{k})^r=\sum_{J=r}^{J=\infty}\frac{p_s^{2J}}{m_o^{2J-r}c^{2J-2r}}G_{Jr}(i-1)
\end{equation}
 , where $G_{Jr}(i-1)$ is defined by the following relation
\begin{equation}
  G_{jr}(i-1)=\sum_{j_1+...j_r=J}g_{j_{1}}(i-1)...g_{j_{r}}(i-1)
\end{equation}
\footnote{Functions like $ G_{jr}$ and later $ F_{jr}(x)$ will not
be named, they are just a substitution to make the equations more
compact. These functions have nothing to do with tensors, and in
general there is no use of any tensors in this work} Substituting
(22) into (19) will give
\begin{eqnarray}
  E_i=\frac{1}{2}\Bigg[\sum_{j_0=1}^{j_0=\infty}\sum_{J=0}^{J=\infty}
  \sum_{r=co.}^{r=J}\frac{p_{s}^{2(j_0+J)}}{m_o^{2(j_0+J)-1}
  c^{2(j_0+J)-2}}[2i-1]^{j_0}2^{j_0}
  \\\nonumber C(1/2,j_0) C(-2j_0+1,r)G_{Jr}(i-1)\Bigg]
\end{eqnarray}
, where $r=co.$ means that the summation will begin from $r=0$ if
$J=0$ and begins from $r=1$ for $J\geq 1$. By substituting
\begin{equation}
  j=J+j_0
\end{equation}
into (24), and using (16) the expression of $K$ will be
\begin{eqnarray}
K(i)=\sum_{l=1}^{l=i}E_l=\sum_{j=1}^{j=\infty}g_j(i)\frac{p_s^{2j}}{m_o^{2j-1}c^{2j-2}}=
\sum_{l=1}^{l=i}\sum_{j=1}^{j=\infty}\frac{p_{s}^{2j}}{m_o^{2j-1}c^{2j-2}}
\sum_{J=0}^{J=j-1}\sum_{r=co.}^{r=J}\\\nonumber
C(1/2,j-J)C(-2j+2J+1,r)[2l-1]^{j-J}2^{j-J-1}G_{Jr}(l-1)
\end{eqnarray}
\par Equating the coefficients of $p_s^j$  in both sides of (26)
gives
\begin{equation}
  g_j(i)=\sum_{l=1}^{l=i}\sum_{J=0}^{J=j-1}\sum_{r=co.}^{r=J}C(1/2,j-J)C(-2j+2J+1,r)[2l-1]^{j-J}2^{j-J-1}G_{Jr}(l-1)
\end{equation}
It is easy to see that any $g_k(l-1)$ that appears on the right
hand side of (27) with order $k<j$, therefore $ g_j(i)$ will be
expressed in terms of $g's$ with lower orders. The calculations
will involve the values of the binomial coefficients, and it also
involves summations like $\sum_{l=1}^{l=1}l^n \hspace{4mm}
(n=0,1,2...)$, that demand using rules concerning the values of a
series of integers in which Bernoulli numbers are used. No details
will be mentioned here about these things since they are available
in many calculus books , like [8], [9].
\par Equation (27) gives
\begin{eqnarray}
g_1(i)=\frac{i^2}{2}\\g_2(i)=-\frac{i^4}{8}-\frac{i^3}{6}+\frac{i}{24}\\g_3(i)=\frac{i^6}{16}+
\frac{11i^5}{60}+\frac{i^4}{8}-\frac{i^3}{16}-\frac{i^2}{16}+\frac{i}{240}
\\g_4(i)=-\frac{5i^8}{128}-\frac{103i^7}{560}-\frac{13i^6}{48}-\frac{37i^5}{960}+\frac{19i^4}{96}+\frac{13i^3}{160}
-\frac{17i^2}{384}-\frac{101i}{6720}\\g_5(i)=\frac{7i^{10}}{256}+
\frac{1823i^9}{10080}+\frac{1219i^8}{2880}+\frac{4079i^7}{13440}-\frac{1607i^6}{5760}-\frac{277i^5}{640}\\
\nonumber +\frac{479i^4}{11520}
+\frac{1957i^3}{10080}+\frac{i^2}{180}-\frac{181i}{6720}\\g_6(i)=-\frac{21i^{12}}{1024}-\frac{31373i^{11}}{177408}-
\frac{2589i^{10}}{4480}-\frac{81199i^9}{107520}+\frac{2627i^8}{26880}+\frac{29983i^7}{26880}\\\nonumber
+\frac{1333i^6}{3072}-\frac{124361i^5}{161280}-\frac{19829i^4}{53760}+\frac{32353i^3}{107520}+\frac{963i^2}{8960}-\frac{12223i}{295680}
\\g_7(i)=\frac{33i^{14}}{2048}+\frac{265537i^{13}}{1537536}+\frac{1182109i^{12}}{1612800}+\frac{29978387i^{11}}{21288960}
+\frac{280981 i^{10}}{460800}\\\nonumber -\frac{748987
i^9}{387072}-\frac{2292077i^8}{1075200}+\frac{5939i^7}{3840}+\frac{963827i^6}{403200}-\frac{2089657i^5}{1935360}\\\nonumber-
\frac{705691i^4}{460800}+\frac{10031797i^3}{21288960}+\frac{57733i^2}{134400}-\frac{175143i}{2562560}\\g_8(i)=-\frac{429i^{16}}
{32768}-\frac{444679i^{15}}{2635776}-\frac{94478431i^{14}}{106444800}-\frac{12578196061i^{13}}{5535129600}-
\\\nonumber \frac{1372584929i^{12}}{638668800}+\frac{164377183i^{11}}{70963200}+\frac{1359253633i^{10}}{232243200}
-\frac{44558251i^9}{38707200}\\\nonumber-\frac{164613529i^8}{19353600}+\frac{45117479i^7}{38707200}+\frac{551387449i^6}{58060800}
-\frac{15167567i^5}{11827200}\\\nonumber-\frac{33346055137i^4}{5109350400}+\frac{3617175179i^3}{5535129600}+\frac{3585115i^2}
{1892352}-\frac{3033083i}{30750720}
\end{eqnarray}
\par To see what these calculation may lead to, a digression would
 be useful at this point. Let us consider a sufficiently high value of
 $i$ such that in all the equations from (28) to (35) only the terms
with higher power of $i$ will be dominating and the other terms
can be neglected, then from (16) and knowing that $p=ip_s$ , the
expression of $K(p)$ will be
\begin{eqnarray}
K(p)\approx\frac{p^2}{2m_o}-\frac{p^4}{8m_o^3c^2}+\frac{p^6}{16m_o^5c^4}-\frac{5p^8}{128m_o^7c^6}
+\frac{7p^{10}}{256m_o^9c^8}-\\\nonumber
\frac{21p^{12}}{1024m_o^{11}c^{10}}+\frac{33p^{14}}{2048m_o^{13}c^{12}}-\frac{429p^{16}}{32768m_o^{15}c^{14}}
+...
\end{eqnarray}
The generating function for the above series is
\begin{equation}
  K(p)\approx\sqrt{m_o^2c^4+p^2c^2}-m_oc^2
\end{equation}

This is nothing but the kinetic energy that was expressed from SR
by using Lorentz transformations.
\par To put things in more specific form, it is useful to write $K$
as a power series of functions
\begin{equation}
 K(p)=\sum_{j=0}^{\infty}
\frac{p_{s}^{j}}{m_{o}^{j-1}c^{j-2}}f_{j}(\frac{p}{m_oc})=\sum_{j=0}^{\infty}
\frac{p_{s}^{j}}{m_{o}^{j-1}c^{j-2}}f_{j}(x)
\end{equation}
\par, where
\begin{equation}
  x=\frac{p}{m_oc}
\end{equation}
 By comparing (38) with (16) and by using the equations from (28) to
(35), the functions $f_j$ can be expressed as power series, which
are
\begin{eqnarray}
f_0=\frac{x^2}{2}-\frac{x^4}{8}+\frac{x^6}{16}-\frac{5x^8}{128}+
\frac{7x^{10}}{256}-\frac{21x^{12}}{1024}+\frac{33x^{14}}{2048}-\frac{429x^{16}}{32768}+...
\\f_1=-\frac{x^3}{6}+\frac{11x^5}{60}-\frac{103x^7}{560}+\frac{1823x^9}{10080}-
\frac{31373x^{11}}{177408}+\frac{265537x^{13}}{1537536}\\\nonumber-\frac{444679x^{15}}{2635776}+...\\
f_2=\frac{x^4}{8}-\frac{13x^6}{48}+\frac{1219x^8}{2880}-\frac{2589x^{10}}{4480}+\frac{1182109x^{12}}{1612800}
-\frac{94478431x^{14}}{106444800}\\\nonumber+\frac{244535971x^{16}}{234823680}+...\\
f_3=\frac{x}{24}-\frac{x^3}{16}-\frac{37x^5}{960}+\frac{4079x^7}{13440}-\frac{81199x^9}{107520}+
\frac{29978387x^{11}}{21288960}\\\nonumber-\frac{12578196061x^{13}}{5535129600}+\frac{4127110357x^{15}}{1230028800}+...\\
f_4=-\frac{x^2}{16}+\frac{19x^4}{96}-\frac{1607x^6}{5760}+\frac{2627x^8}{26880}+\frac{280981x^{10}}{460800}
-\frac{1372584929x^{12}}{638668800}\\\nonumber
+\frac{225909079003x^{14}}{46495088640}-\frac{156769939621x^{16}}{17220403200}+...
\end{eqnarray}
\section{The Series Solution}
\par The generating function must be found in order to explore the
dispersion relation corrected to the $j$ order, that is because
one of the aims of this work is to calculate corrections to EDR
and find the effect of these corrections on the velocity. Finding
the form of the power series of $f_j$ will lead to guess only
$f_0$ which represents the kinetic energy term in EDR , and $f_1$
, which is the first order correction. They have the following
expressions
\begin{equation}
f_{0}(x)=\sqrt{1+x^2}-1
\end{equation}
\begin{equation}
  f_{1}(x)=\frac{tan^{-1}(x)-x}{2\sqrt{1+x^2}}
\end{equation}
 The other series of $f_2, f_3,...$ are too complicated to give
any clue about their generating functions. Therefore it is
important to find another technique that can express the functions
$f_j$ directly. The following section will discuss such a
technique that involves solving first order differential
equations. The equations will become increasingly long and
complicated with the increase of the order of correction.
According to my experience the series solution is vital for
checking the results.
\section{The Generating Function Solution}
The change of  energy of a particle due to  $(i+1)p_s$ pulse can
be written in terms of the total energy of the particle when
$p=ip_s$ . This can be done by using (8), and (9) that give
\begin{equation}
  E_{i+1}(p+p_s)=\frac{1}{2}\left(-E^{(t)}(p)+
\sqrt{(E^{(t)}(p))^2+2p_{s}^2c^2(2i+1)} \right)
\end{equation}
By using a Taylor expansion it is easy to see that
\begin{equation}
  E_{i+1}(p+p_s)=E^{(t)}(p+p_s)-E^{(t)}(p)=(e^{p_sD}-1)E^{(t)}(p)
\end{equation}
where the operator $D$ is defined by the following equation
\begin{equation}
  D=\frac{\partial}{\partial p}=\frac{1}{m_oc}\frac{\partial}{\partial
  x}\equiv \frac{1}{m_oc}\partial
\end{equation}
From (47), (48), (49) ,and (38), and after some elaboration, a
useful expression can be reached, that is
\begin{eqnarray}
\sum_{l=1}^{l=\infty}\frac{p_s^l}{m_o^lc^l}\Bigg[\sum_{n+m+j+k=l
}\frac{p_s}{m_oc}\frac{\partial n f_j(x)}{n!}\frac{\partial
mf_k(x)
}{m!}+\sum_{n+j=l}\frac{\partial nf_j(x)}{n!}\\
\nonumber  + \sum_{j+n+k=l} f_k(x)\frac{\partial
nf_j(x)}{n!}\Bigg] =\frac{2xm_oc+p_s}{2m_oc}\frac{p_s}{m_oc}
\end{eqnarray}
 , where $n,m\geq 1$ . Taking the coefficients of $p_s/m_oc$ on
both sides of (50) will give a first order differential equation
for $f_0$, that is
\begin{equation}
(1+f_0)\frac{\partial f_0}{\partial x}=x
\end{equation}
Solving the above equation after applying the condition $f_0(0)=0$
gives exactly  the same expression of (45) for $f_0$. To express
$f_1$ the coefficients of $p_s^2/m_o^2c^2$ must be equated on both
sides of (50) to get
\begin{equation}
  \left(\frac{\partial f_0}{\partial x}\right)^2
  +\left(1+f_0\right)\left(\frac{1}{2}\frac{\partial 2f_0}{\partial
  x^2}+\frac{\partial f_1}{\partial
  x}\right)+f_1\left(\frac{\partial f_0}{\partial
  x} \right)=\frac{1}{2}
\end{equation}
Substituting for $f_0$ from (45) and after that solving the
resultant first order differential equation under the condition
$f_1(0)=0$ will give exactly the expression of $f_1$ in (46). The
process can be continued in similar way to find $f_2,f_3,f_4,..$.
Because of the increasing complexity, there will be no mention
here of the details of calculating $f_2,f_3,f_4$ . The results can
be summarized as follows
\begin{equation}
f_{0}(x)=\sqrt{1+x^2}-1
\end{equation}
\begin{equation}
  f_{1}(x)=\frac{\tan^{-1}(x)-x}{2\sqrt{1+x^2}}
\end{equation}
\begin{equation}
f_{2}(x)=\frac{x^2(\tan^{-1}( x))^2}{8(1+x^2)^{\frac{3}{2}}}
\end{equation}
\begin{eqnarray}
f_{3}(x)=-\frac{1}{192}(1+x^2)^{\frac{-5}{2}}\Big(-5x-3x^3-
(3+6x+3x^2)\tan^{-1}(x)\\\nonumber+12x^3(\tan^{-1}(x))^2+
12x^2(\tan^{-1}(x))^3 \Big)
\end{eqnarray}
\begin{eqnarray}
  f_{4}(x)=-\frac{1}{1152}(1+x^2)^{\frac{-7}{2}}\Big(9x^2-3x^4-4x^6+(54x+33x^3+9x^5)\\\nonumber\tan^{-1}(x)+(9+18x^2-27x^4)
  (\tan^{-1}(x))^2-\\\nonumber96x^3(\tan^{-1}(x))^3+(9x^4-36x^2)(\tan^{-1}(x))^4\Big)
\end{eqnarray}
To check whether these results are correct or not, the Maclaurin
expansion should be found for each of the $f_j$ functions written
in the above equation. If the expansion coincide with the
corresponding one in (40) to (44) then this will certify that the
results are correct.
\section{The Convergence of the Series of Functions}
\par The series of $ E^{(t)}(x)$ can be  written as
\begin{equation}
  E^{(t)}(x)=m_oc^2+m_oc^2f_0(x)+\sum_{j=1}^{j=\infty}u_j(x)
\end{equation}
, where the third term on the left hand side is a series of
functions with
 \begin{equation}
 u_j(x)=\frac{p_s^j}{m_o^{j-1}c^{j-2}}f_j(x)
\end{equation}
In this section, the conditions for the convergence of the series
$\sum u_j(x)$ will be discussed.
\par The expression given for the functions $f_1,...,f_4$  in
equations from (54) to (57) give no clue about a general rule that
can be reached by mathematical induction to express $f_j$ for any
value of $j$. The Weierstrass theorem will be useful to verify the
convergence for a case when the general term of the series is
unknown [10],[11]. The state of the theorem is: Suppose $\{u_n\}$
is a sequence of functions defined on $ E$, and suppose
\begin{equation}
  \left|u_j(x)\right|\leq M_j ,\qquad(x\in E,j=1,2,3...)
\end{equation}
then $\sum u_j(x)$ converges uniformly on $E$ if $\sum M_j$
converges. For this case $E$ is the closed interval  $[0,\infty]$.
By plotting the functions $f_1,...,f_4$ it is obvious that $|f_j|$
are bounded on the interval $[0,\infty]$. The graphs give
 \begin{eqnarray}
\sup_{x \in [0, \infty ]}|f_1(x)|=0.5\\\sup_{x \in [0, \infty
]}|f_2(x)|=0.0565\\\sup_{x \in [0, \infty
]}|f_3(x)|=0.0159\\\sup_{x \in [0, \infty ]}|f_4(x)|=0.0073
 \end{eqnarray}
 Equations from (61) to (64) together with (59) and (60) suggest more than one
 form of $\{M_j\}$, such as:
\begin{eqnarray}
M_j=\frac{1}{2}\sum_{j=1}^{j=\infty}\frac{p_s^j}{m_o^{j-1}c^{j-2}}\\
M_j=\frac{1}{2}\sum_{j=1}^{j=\infty}\frac{p_s^j}{m_o^{j-1}c^{j-2}j}
\end{eqnarray}
\par The $\rho$ -test for both two expressions of (65) and (66) will give
convergence under the condition
\begin{equation}
  m_oc>p_s
\end{equation}
, which means that the suggested mathematical treatment by series
will give the correct result only if $ m_oc>p_s$.
 One might say that the expression
\begin{equation}
M_j=\frac{1}{2}\sum_{j=1}^{j=\infty}\frac{p_s^j}{m_o^{j-1}c^{j-2}2^{j-1}}
\end{equation}
 also satisfies (60), and gives a more flexible result since it
 leads
to the condition  $ m_oc>p_s/2$, but this wouldn't be safe.
Adopting such a result demands the knowledge of at least $f_5$ and
$f_6$ in order to conform that (60) will be satisfied. Finding
these functions demand doing a very long and tedious calculations.
To calculate the dispersion relation for $ m_oc<p_s$, the
mathematical approach must be modified, that will not be discussed
in this work, only for special case when the momentum is few
$p_s$'s, that is when the velocity of the particle will be studied
next section.
\section{The Velocity of Particles}
\par The relation between the velocity and momentum is important, it will be used to explain why the velocity of light
 seems to be independent on the velocity of the source, and under what conditions this will be changed.
 \par The relation between the velocity and total energy of a particle
will allow  us to check whether the present treatment will lead to
the same Einstein relation between kinetic mass, rest mass and
velocity. It will be shown that this relation will be expressed as
a zeroth approximation.
 \subsection{The Velocity as a Function of $p$} By definition, a velocity of a particle is related to it's
linear momentum by the following relation
\begin{equation}
  \overrightarrow{v}=\frac{\overrightarrow{p}}{m}
\end{equation}
or
\begin{equation}
  v=\frac{p}{m}
\end{equation}
 where
\begin{equation}
  m=\frac{E^{(t)}}{c^2}=m_o+\frac{K}{c^2}
\end{equation}
From (70) and (71), (38) can be written as
\begin{equation}
  v=\frac{p}{m_o+m_of_0(x)+\sum_{j=1}^{j=\infty}\frac{p^j}{m_o^{j-1}c^j}f_j(x)}
\end{equation}
Applying the binomial theorem to the term at the denominator in
(72) gives
\begin{eqnarray}
  v=\frac{p}{m_o+m_of_0}\sum_{r=0}^{r=\infty}\frac{1}{(m_oc+m_of_0)^r}\;C(-1,r)\\\nonumber
  \sum_{j_1=1}^{j_1=\infty}...\sum_{j_r=1}^{j_r=\infty}\frac{p_s^{j_1+...j_r}}{m_o^{j_1+...j_r-r}c^{j_1+...j_r}}f_{j_1}...f_{j_r}
\end{eqnarray}
After substituting $j_1+...j_r=j$ , (73) can be written as
\begin{equation}
v=\sum_{j=0}^{j=\infty}\sum_{r=co.}^{r=j}\frac{p_s^j}{m_o^jc^{j-1}}\;C(-1,r)F_{jr}(x)
\end{equation}
, where
\begin{equation}
F_{jr}(x)=\frac{x}{(1+f_0(x))^{r+1}}\sum_{j_1+...j_r=j}f_{j_1}(x)...f_{j_r}(x)
\end{equation}
To get an iterated value for the velocity, $v$ will be written as
\begin{equation}
  v(x)=\sum_{j=0}^{j=\infty}\frac{p_s^j}{m_o^jc^j}v_j(x)
\end{equation}
, where $v_0$ is the velocity corrected to the zero order, $v_1$
is the first order correction to the velocity as a function of
momentum, $v_2$ is the second order correction,... and so forth.
 From (76), (74), the expression for $v_j$ can be found, which is
\begin{equation}
  v_j(x)=\sum_{r=co.}^{r=j}c \;C(-1,r)F_{jr}(x)
\end{equation}
\subsection{The velocity of Light}
\par To see what will be the velocity for a particle with $p\gg
m_oc$, the functions $F_{jr}(x)$ must be calculated , and then
find $v_j$ for $j=0,1,2,..$. Using (75), (53),(54) and (55) give
\begin{eqnarray}
F_{00}=\frac{x}{(1+x^2)^{\frac{1}{2}}}\qquad
F_{11}=\frac{x(\tan^{-1}(x)-x)}{2(1+x^2)^{\frac{3}{2}}}\\\nonumber
F_{21}=\frac{x^3(\tan^{-1}(x))^2}{8(1+x^2)^{\frac{5}{2}}}\qquad
F_{22}=\frac{x(\tan^{-1}(x)-x)^2}{4(1+x^2)^{\frac{5}{2}}}\\\nonumber...etc.
\end{eqnarray}
For the case $p\gg m_oc$ (or $x\rightarrow\infty$) it is obvious
from (78) that all the functions $F_{jr}\rightarrow 0$, except the
function $F_{00}\rightarrow 1$. Therefore (77) will give
$v_j\rightarrow 0$ for $j=1,2,...$ and $v_0\rightarrow c$ , and
directly (74) will give
\begin{equation}
  \lim_{x\rightarrow\infty}v=c
\end{equation}
that coincides with experiment.
\par To prove that the velocity of light is independent of the
velocity of the source a simple fact will be used. If the source
moved away or approached the observer, the momentum of the photon
will change, that is decrease in the first case and increase in
the second case. This fact is applied not only to the emission of
photons but to the emission of any particle. Performing any
experiment to measure the effect of the velocity of the source on
the velocity of light will not give any indication that the
velocity of light is changing as long as the condition $p\gg m_oc$
does not change. To see why, the derivative of $v$ with respect to
$p$ will be found, it is
\begin{equation}
\frac{d v}{d p}=\frac{d v}{d x}\frac{d x}{d
p}=\sum_{j=0}^{j=\infty}\sum_{r=co.}^{r=j}\frac{1}{m_o}
C(-1,r)\frac{dF_{jr}(x)}{dx}
\end{equation}
From (78) and by induction it can be proved that
\begin{equation}
  \lim_{x\rightarrow\infty}\frac{dF_{jr}(x)}{dx}=0
\end{equation}
which means that there will be no change in the velocity as a
result of a change in momentum as long as $p\gg m_oc$. More
precisely the change will be so small that it is beyond detection.
 This is a result and not a postulate as in the case of SR.
\par In this work the difference is that, in principle, there could be
an experiment that measures a change in velocity of photons, but
this experiment must involve a source moving away from the
observer with extremely high velocity to produce a high Doppler
shift for photons that have originally a very long wave length,
such that $x$ will not be big enough to make the right hand side
of (80) approaching zero. Then the change of the velocity can be
detected. No further specifications can be given about such an
experiment since the rest mass of the photon is still unknown.

\subsection{The Total Energy Velocity Relation} \par To find the
relation between total energy and velocity, (9) will be written as
\begin{equation}
  E^{(t)}=K+m_oc^2=m_oc^2+\sum_{j=0}^{j=\infty}\frac{p_s^j}{m_o^{j-1}c^{j-2}}f_j(\frac{E^{(t)}v}{m_oc^3})
\end{equation}
\par The above equation can be reached directly by substituting
\begin{equation}
x=\frac{p}{m_oc}=\frac{1}{m_oc}(\frac{E^{(t)}}{c^2}v)
\end{equation}
The expansion of the velocity as a function of total energy is
\begin{equation}
  v=\sum_{j=0}^{j=\infty}\frac{p_s^j}{m_o^jc^j}\beta_j \;c
\end{equation}
, where
\begin{equation}
  \beta_j=\frac{\tilde{v}_j}{c}\qquad (j=0,1,2...)
\end{equation}
, and $\tilde{v}_j$ is the j-correction of the velocity as a
function of energy. Writing the equations in terms of $\beta$ will
give the equations in a more compact form, as we will see later.
To find a series solution, the following substitution will be made
\begin{equation}
  x=\xi_0+\xi_1+...+\xi_n+...=\xi_0+\sum_{k=1}^{\infty}\xi_k
\end{equation}
,where
\begin{equation}
\xi_{n}=\frac{E^{(t)}\beta_n}{m_oc^2}\frac{p_s^n}{m_o^n c^n}
\end{equation}
 Applying the chain rule and using (86) gives
\begin{equation}
  \frac{\partial}{\partial \xi_k}= \frac{\partial x}{\partial
  \xi_k}\frac{\partial}{\partial x}=\frac{\partial}{\partial x}
\end{equation}
Here it must be noted that for $ x=p_s/mc$ the only solution for
(86) and (87) is
\begin{eqnarray}
\nonumber
x=\xi_0=\frac{E^{(t)}\beta_0}{m_oc^2}\left(\frac{p_s}{m_o
c}\right)^{0}=\frac{E^{(t)}v_0}{m_oc^3}=\frac{p_s}{m_oc}
\end{eqnarray}
 Accordingly the Taylor series of (82) about $\xi_0$ for the several
variables $\xi_1,\xi_2,...\xi_k,...$ will be
\begin{eqnarray}
  E^{(t)}=m_oc^2+m_oc^2f_0(\xi_0)+\sum_{j=1}^{j=\infty}\frac{p_s^j}{m_o^{j-1}c^{j-2}}f_j(\xi_0)+
  \sum_{j=1}^{j=\infty}\frac{p_s^j}{m_o^{j-1}c^{j-2}}\\\nonumber
  \sum_{J=0}^{J=j-1}\sum_{l=1}^{l=j-J}\frac{1}{l!}\left(\frac{\partial l
  f_J}{\partial x^l}\right)_{x=\xi_0}\left(\frac{E^{(t)}}{m_oc^2}\right)^l\left[\sum_{k_1+...+k_l=j-J} \beta_{k_1}...
  \beta_{k_l}\right]
\end{eqnarray}
where the above equation was reached by arranging term with equal
power of $p_s$. Taking the coefficients of $p_s^0$ in (89) gives
\begin{equation}
   E^{(t)}=m_oc^2+m_oc^2f_0(\xi_0)=\sqrt{m_o^2c^4+\frac{ E^{(t)2}v_0^2}{c^2}}
\end{equation}
The above relation gives exactly the Einstein relation between
kinetic mass, rest mass, and velocity:
\begin{equation}
\frac{E^{(t)}}{c^2}=m=\frac{m_o}{\sqrt{1-\frac{v_0^2}{c^2}}}
\end{equation}
Equation (91) gives
\begin{equation}
  v_0=c\;\sqrt{1-\frac{m_o^2c^4}{ E^{(t)2}}}
\end{equation}
Equating the coefficients of equal $p_s^j$ on both sides of (89)
for $j=1,2,...$  gives
\begin{eqnarray}
  \tilde{v}_j=-\left[ \frac{E^{(t)}}{m_0c^3}\left(\frac{\partial f_0}{\partial
  x}\right)_{x=\xi_0}\right]^{-1}\Bigg(f_j(\xi_0)+  \sum_{J=0}^{J=j-1}\sum_{l=1}^{'l=j-J}\frac{1}{l!}\left(\frac{\partial l
  f_J}{\partial x^l}\right)_{x=\xi_0}\\\nonumber\left(\frac{E^{(t)}}{m_oc^2}\right)^l\left[\sum_{k_1+...+k_l=j-J} \beta_{k_1}...
  \beta_{k_l}\right]  \Bigg)
\end{eqnarray}
The prime on the second integral means that the term with $J=0$
and $l=1$ is not included in the summation. The first correction
to the velocity as a function of total energy is
\begin{equation}
  \tilde{v_1}=c\;\left[\frac{2 E^{(t)}}{m_oc^2}\sqrt{\frac{ E^{(t)2}}{m_o^2c^4}-1}\right]^{-1}\left(\sqrt{\frac{ E^{(t)2}}{m_o^2c^4}-1}-tan^{-1}(\sqrt{\frac{ E^{(t)2}}{m_o^2c^4}-1})\right)
\end{equation}
The plot of the above function against $E^{(t)}/m_oc^2$ has a
maximum value  of $\tilde{v}_{1max}\approx 0.1c$ at
$E^{(t)}/m_oc^2\approx 2.2$ and the function rapidly approach zero
as $E^{(t)}/m_oc^2\rightarrow \infty$ .In any event, for particles
with $m_oc\gg p_s$ like electrons and protons the effect of this
term on the value of the velocity will be very small, while for
particles with small masses such that $m_oc \sim p_s$ the
correction has larger impact when $E^{(t)}\sim m_oc^2$.
\subsection{Faster Than Light Transition}One of the important
result of this work is the possibility of having particles that
can travel faster than light, but as will be shown later, these
particles must have a small mass in order to reach such
velocities. Moreover the momentum of these particles must be very
low, so that the particle can exceed the velocity of light in a
measurable value. The calculations in this section will be done by
using exact relations and not approximate relations as it was done
in the previous sections, that is because the energy can be
calculated exactly by using (8) and (9) when the calculation
involves only momenta with few $p_s$'s. It will be shown next how
the velocity changes with the change of mass for a certain value
of the momentum.
\par 1.The case $p=1p_s$:
\begin{equation}
E^{(t)}=\frac{1}{2}\left(m_oc^2+\sqrt{m_o^2c^4+2p_s^2 c^2}\right)
\end{equation}
This gives
\begin{equation}
  v=\frac{p_s}{m}=\frac{2\delta}{1+\sqrt{1+2\delta^2}}c
\end{equation}
, where $\delta=p_s/m_oc$.
\begin{figure}
\begin{center}
\includegraphics[height=11cm,width=6cm,angle=-90]{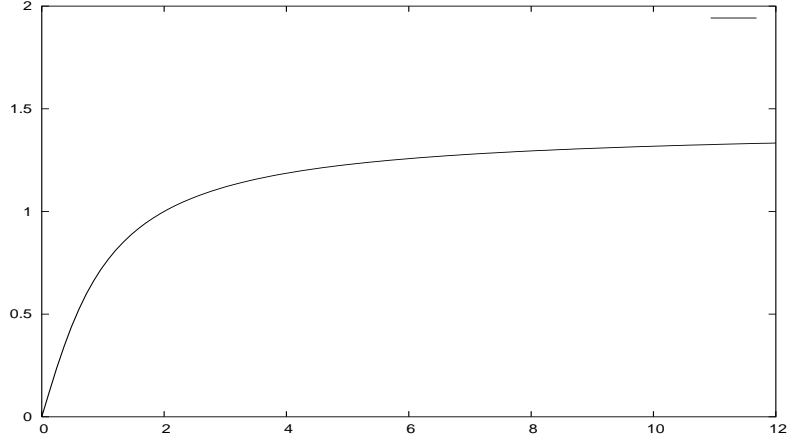}
\end{center}
\caption{The plot of the velocity in the units of $c$
versus $\delta $ for the case $p=p_s$.}
\end{figure}

 It is easy to prove that the above
relation will give $v=c$ for $\delta=2$ which means that a
particle with momentum $p_s$ and mass $m_o=p_s/2c$ will have a
velocity equal to $c$. The velocity will increase with the mass
decrease further until it reaches $v=\sqrt{2}c$, which represent
the upper bound of the velocity for any mass with momentum $p_s$
no matter how small it is.

\par 2.The case $p=2p_s$ :
\begin{equation}
v=8\delta
c\left[1+\sqrt{1+2\delta^2}+\sqrt{2+26\delta^2+2\sqrt{1+2\delta^2}}\right]^{-1}
\end{equation}
\begin{figure}
\begin{center}
\includegraphics[height=11cm,width=6cm,angle=-90]{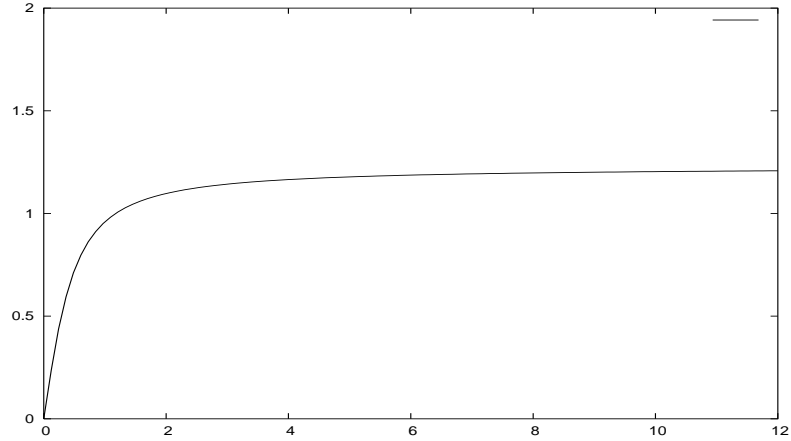}
\end{center}
\caption{The plot of the velocity in the units of $c$
versus $\delta $ for the case $p=2p_s$}
\end{figure}
The plot of $v$ against $\delta$ for this case shows that the
velocity of a particle will increase with the decrease of the mass
until it will reach $c$ at $m_o=p_s/(1.22)c$. It will increase
further with the decrease of $m_o$ (increase of $\delta$) until it
will reach an upper bound when $\delta \rightarrow \infty$, then
\begin{equation}
  v=\frac{8c}{\sqrt{2}(1+\sqrt{13})}\approx 1.23c
\end{equation}
\par3.The case $p=3p_s$:
\begin{eqnarray}
  v=24\delta c\Bigg\{1+\sqrt{1+2\delta^2}+\sqrt{2+26\delta^2+2\sqrt{1+2\delta^2}}+\hspace{20mm}\\\nonumber\sqrt{4+188\delta^2+
  4\sqrt{1+2\delta^2}+2(1+\sqrt{1+2\delta^2})\sqrt{2+26\delta^2+2\sqrt{1+2\delta^2}}}\Bigg\}^{-1}
\end{eqnarray}
The plot(will notbe shpwn here) of $v$ against $\delta$ for this case shows that the
velocity of a particle increases with the decrease of the mass
until it reaches  $c$ at $m_o=p_s/(0.89)c$. It increases further
with the decrease of $m_o$ (increase of $\delta$) until it reaches
 an upper bound when $\delta\rightarrow \infty$. Then
\begin{equation}
  \lim_{\delta\rightarrow \infty}v=\frac{24c}{\sqrt{2}+\sqrt{26}+\sqrt{188+2\sqrt{2}\sqrt{26}}}\approx
  1.157 c
\end{equation}
 It is obvious that the value  $\lim_{\delta\rightarrow \infty}v$
 decreases
with the increase of $p$ , and it will be equal to $c$ for
$p_s/m_0c \rightarrow \infty$, which means that the value that a
particle can exceed the velocity of light is getting smaller and
smaller with the increase of $p$. Therefore the faster than light
transition will be discernable only at very low values of $p$, and
when the particles involved have very small masses. For the first
case the mass must be smaller than $p_s/2c$ for the second case it
must be smaller than $p_s/(1.22)c$ and for the third case it must
be smaller than $p_s/(0.89)c$.
\section{Summary of Conclusions}
This work has proved that it is possible to get the Einstein
dispersion relation without using Lorenz transformations. The EDR
appears as a zeroth approximation. The plot of the functions $f_0,
f_1, f_2, f_3, f_4$ against $x$ shows that $lim_{x\rightarrow
\infty}f_0=\infty$ , and $lim_{x \rightarrow \infty} f_1(x)=1/2$
while $\lim_{x \rightarrow \infty }f_k(x)=0 $ for $k=2,3...$. This
means that the EDR will be more and more accurate with the
increase of momentum. The derived dispersion relation agrees with
the one of SR for high energies. But for low energies and masses
this will not be the case because these terms may have a
considerable value. One of the difficulties here is the unknown
value of the universal constant $p_s$. Because of that it is not
possible to predict at what low energies the corrections are
needed.
\par The first assumption of SR that the velocity of light is
independent of the velocity of the source is a result here that
could be concluded. This new understanding lead to an experimental
conditions in which the first assumption of SR can be violated.
The violation can occur when the wave length of the photons is
very long and the source is moving away from the observer with
very high velocity.
\par Another important result is it that is  permissible for
nonzero mass particles to travel with a velocity of $c$ and
faster. This will solve the conceptual difficulty of zero mass
photon since it is permitted to all particles to have a rest mass
here including photons.
\par The faster than light transition is permitted here only under
a condition that the masses of the particles must have a certain
degree of smallness compared to $p_s/c$. If there are no such
particles in nature (including photons) then there will be no FTL.
On the other hand, if FTL is detected then it is the proof of an
existence of such particles with such small rest masses. The
results of the previous section show that the there will be no
hope of detecting such transition for photons with short wave
length (high momentum) because from (77) such a photon will travel
with velocity very close to $c$, the FTL could be detected if the
experiment is designed to measure the velocity of photons with
extremely long wave length. It has been shown that the largest
possible velocity of a particle is $\sqrt{2}c$ that is when
$p=p_s$ and the mass approaches zero.
\par Here I must admit that there is no usefulness of the new approach in
applied physics unless faster than light transition is detected
and measured experimentally, it would be hard to justify the
replacement of one postulate of SR with three postulate if FLT
will not be detected.
\section*{References}
[1] R. P. Feynman, F. B. Morinigo, W. G.Wagner, B. Hatfield,
 J. Preskill,  \& K. S. Thorne,  Feynman
 Lectures on Gravitation,( Addison-Wesely Publishing
 Company, 1995).\newline
[2] H. P. Schwartz, Introduction to Special Relativity,
(McGraw-Hill, 1968.
\newline[3] H. Poincar\'{e}, Arch. neerland. Sci vol2, 5, p252
(1900)
\newline[4] F. Hasen\"{o}hrl, Ann. Physik vol.16, 4 , p593(1905)
\newline[5] Cunningham, The Principle of Relativity (Cambridge
University Press, 1914).
\newline[6] M. Jammer, Concepts of Mass in Classical and Modern
 Physics, ( Dover Publications,Inc., 1997)
\newline[7] Al-Hashimi,M. H. K. , Discrete Momentum Mechanics and
Collision of Particles, under preparation.
 \newline[8] K. F. Riley,  M. P. Hobson, \& S. J. Bence,
 , Mathematical Methods for Physics and Engineering, (Cambridge
 University Press, 2002).
 \newline[9] J. Spanier, \&  K. B. Oldham, An  Atlas of
 functions, ( Washington a.o.: Hemisphere Publishing Corporation; Berlin a.o.:
 Springer, 1987).
\newline[10] W. Rudin, Principle of Mathematical Analysis, (McGraw-Hill, 1976).
\newline[11] Y. S. Bugrov, S. M. Nikolsky, Higher
Mathematics, ( Mir Publisher, 1982).
\end{document}